\documentclass[preprint,showpacs,preprintnumbers,amsmath,amssymb,endfloa
ts*]{revtex4}
\usepackage{graphicx}

\begin{document}

\thispagestyle{empty}
\title{
Thermal corrections in
Casimir interaction between metal
and dielectric
}
\author{
B.~Geyer,\footnote{
E-mail: geyer@itp.uni-leipzig.de}
G.~L.~Klimchitskaya,\footnote{
On leave from
North-West Technical University,
St.Petersburg, Russia. \\
E-mail: Galina.Klimchitskaya@itp.uni-leipzig.de}
and V.~M.~Mostepanenko\footnote{
On leave from
Noncommercial Partnership ``Scientific Instruments'' Moscow, Russia.\\
E-mail: Vladimir.Mostepanenko@itp.uni-leipzig.de}
}

\affiliation{
Center of Theoretical Studies and Institute for Theoretical Physics,\\
Leipzig University, Augustusplatz 10/11, D-04109 Leipzig, Germany 
}

\begin{abstract}
The Casimir interaction between two thick parallel plates,
one made of metal and the other of dielectric, is investigated at
nonzero temperature. It is shown that in some temperature
intervals the Casimir pressure and the free energy of a fluctuating 
field are the nonmonotonous
functions of temperature and the respective Casimir entropy can
be negative. The physical interpretation for these conclusions
is given. At the same time we demonstrate that the entropy vanishes
when the temperature goes to zero, i.e., in the Casimir interaction
between metal and dielectric the Nernst heat theorem is satisfied.
The investigation is performed both analytically, by using the  
model of an ideal metal and dilute dielectric or dielectric with a
frequency-independent dielectric permittivity, 
and numerically for real metal (Au)
and dielectrics with different behavior of the dielectric permittivity
along the imaginary frequency axis (Si and $\alpha\mbox{-Al}_2\mbox{O}_3$).
\end{abstract}
\maketitle

\section{Introduction}

Recently, there has been a renewed interest in the
investigation of the dispersion
forces which originate from the existence of zero-point electromagnetic
oscillations. The most well known physical phenomena of this kind are
the van der Waals and Casimir forces (see, e.g., Refs.~\cite{1,2,3,4}).
Both forces are of the same nature and are related to nonrelativistic
and relativistic situation, respectively. In the last few years many
measurements of the Casimir force between metals were performed
\cite{5,6,7,8,9,10,11,12,13,14}.
The increasing interest in experimental research of dispersion forces
is motivated by their role in nanotechnology \cite{15,16,16a}
and for constraining predictions of modern theories of fundamental
interactions \cite{13,14,17,18,19} (see also Ref.~\cite{20} covering
both the experimental and theoretical developments on the subject).
Quite recently both the van der Waals and Casimir forces were shown
to be important \cite{21,22,23} in experiments on quantum reflection
and Bose-Einstein condensation.

The theoretical description of the Casimir force between real metals
at nonzero temperature is problematic. The direct 
application
\cite{24,25,26,27} of the Lifshitz formula \cite{Lifshitz} 
in combination
with the dielectric function of the Drude model leads \cite{28}
to a violation of Nernst's heat theorem in the case of perfect
crystals, i.e., to a nonzero value of the Casimir entropy at zero
temperature depending on the separation distance between the plates.
As was shown in Refs.~\cite{27,29}, for crystals with defects or
impurities Nernst's heat theorem is satisfied, so that the entropy
of a fluctuating field is equal to zero at zero temperature.
This, however, does not solve the problem since a perfect crystal is
a truly equilibrium system with non-degenerate dynamic state of
lowest energy. According to quantum statistical physics, the Nernst
heat theorem is valid for any system possessing this property.
It follows that any formalism applied to a perfect crystal must satisfy
the Nernst heat theorem, whereas the approach of Refs.~\cite{24,25,26,27}
does not. It is notable that this approach predicts {\em large} thermal
corrections to the Casimir force at short separations as compared with
the case of ideal metals.

Other theoretical descriptions of the thermal Casimir force between 
real metals are based on the use of the Lifshitz formula combined with
the plasma model dielectric function \cite{30,31,32} or with the
Leontovich surface impedance \cite{28,33}. In the framework of these
approaches the Nernst heat theorem is satisfied for both perfect
lattice as well as lattice with some small concentration of impurities.
They predict {\em small} thermal corrections to the Casimir force at short
separations in qualitative agreement with the case of ideal metals.

Recent experiments reveal the possibility to test the validity of
different theoretical approaches to the thermal Casimir force.
Thus, the first modern measurement of the Casimir force by means
of a torsion pendulum \cite{5} was found \cite{34,35} to be in
disagreement with Refs.~\cite{24,25,26,27} and consistent with
Refs.~\cite{28,30,31,32,33}. The most precise and accurate modern
experiment by means of a micromechanical torsional oscillator 
\cite{13,14} excludes
the theoretical approach of Refs.~\cite{24,25,26,27} at 99\%
confidence and is consistent with the theoretical approaches of
Refs.~\cite{28,30,31,32,33}.
Notice that these conclusions concerning the comparison between
theory and experiment are not universally accepted (see the
discussion on this subject in Refs.~\cite{36,37}).

A distinguishing feature of the theoretical approach of 
Refs.~\cite{24,25,26,27} is the nonmonotonous dependence of the
Casimir pressure and the free energy of a fluctuating field 
on the temperature.
There are some temperature intervals where the magnitudes of the
Casimir pressure and the free energy decrease with the increase
of temperature. As a result, there are temperature intervals
where the entropy
of the fluctuating field takes negative values.
The same is valid for hypothetical (nonexistent) dielectrics
with a frequency-independent dielectric permittivity
$\varepsilon=100$ or higher \cite{26}.
By contrast, in the approaches of Refs.~\cite{28,30,31,32,33} for
metals and also for some real dielectrics considered in 
Ref.~\cite{28}, the magnitudes of the Casimir free energy and
pressure increase with temperature. There is, however, no general
proof that the magnitudes of the Casimir free energy and pressure
must increase with temperature for all real materials.

As can be seen from Ref.~\cite{21}, the magnitudes of the 
Casimir-Polder free energy and force for the case of an atom interacting
with a metal wall may decrease within some temperature region.
The respective entropy of the fluctuating field becomes negative.
When taken into account that the Lifshitz formula for an atom
near a metal wall is obtained \cite{21,38} from the case of
a rarefied dielectric wall spaced parallel to a metal wall,
it may be supposed that the primary reasons for the nonmonotonous
behavior of the free energy are inherent in the Casimir interaction
between metal and dielectric.

In this paper we investigate the Casimir pressure, free energy 
and entropy in
a configuration of two thick parallel plates, one being metallic and 
the other dielectric. Previously, only two metallic or two dielectric
plates were considered in the retarded regime (see also 
Refs.~\cite{39,40} where a so-called ``unusual pair of plates''
was treated -- one of which is made of an ideal metal and the other is
infinitely permeable). In the nonretarded limit of the van der Waals
force the Hamaker constants for the configuration of a metallic and a
dielectric plate were calculated in Ref.~\cite{41}.
On the basis of the Lifshitz formula we find the analytic expressions
for the Casimir pressure, free energy and entropy in the 
configuration of a plate
made of ideal metal and a parallel plate made of dilute dielectric
separated by a gap of thickness $a$. It is shown that there are
intervals on the temperature axis where the magnitude of the Casimir
pressure decreases with the increase of temperature, whereas
the magnitude of the free energy increases with the temperature
and the Casimir entropy is positive. The asymptotic expressions
of the obtained exact formulas at both low and high temperatures are
presented and the validity of the Nernst heat theorem is demonstrated.
The possibility of a decreasing Casimir energy and of negative values of the 
entropy of a fluctuating field within some temperature intervals is 
demonstrated for the cases of ideal and real (Au) metals near
dielectrics with arbitrary constant $\varepsilon$ (not dilute) and
real dielectrics with the frequency-dependent dielectric permittivities
(Si, $\alpha-\mbox{Al}_2\mbox{O}_3$).  The physical interpretation of
that negativeness of entropy is discussed. From this we give a
positive answer to the above question of whether the entropy of
a fluctuating field in the Casimir regime can be negative. This is
important not only for the theory of the thermal Casimir force but
also for experimental investigations of the Casimir interaction 
between metal and semiconductor \cite{42}.

The paper is organized as follows. In Sec.~II the main notation is
introduced and the Casimir free energy, entropy and pressure are
found analytically in the framework of the exactly solvable model
(one plate is made of ideal metal and the other of dilute dielectric).
Sec.~III contains the results of analytical and
numerical computations performed for
the case of ideal-metal plate spaced parallel to the plate of
dielectric with some frequency-independent $\varepsilon$ (not dilute).
The case of Au plate and real dielectrics with the frequency-dependent
dielectric permittivities (Si, $\alpha-\mbox{Al}_2\mbox{O}_3$) is
considered in Sec.~IV. Sec.~V contains our conclusions and
discussion.

\section{Casimir interaction between an ideal metal and a dilute
dielectric}

We consider the configuration of two thick parallel plates, one made
of a real metal and another of dielectric at a separation $a$ and
temperature $T$ in thermal equilibrium. The Casimir free energy per
unit area is given by the Lifshitz formula \cite{Lifshitz,38}.
In terms of dimensionless variables it can be represented in the form
\begin{eqnarray}
&&{\cal F}(a,T)=\frac{k_B T}{8\pi a^2}
\sum\limits_{l=0}^{\infty}\left(1-\frac{1}{2}\delta_{l0}\right)
\nonumber \\
&&\phantom{aaa}
\times\int_{\zeta_l}^{\infty}y\,dy\left\{\ln\left[1-
r_{\|}^{\rm M}(\zeta_l,y)r_{\|}^{\rm D}(\zeta_l,y)e^{-y}\right]\right.
\nonumber \\
&&\phantom{aaa}
\left.+\ln\left[1-
r_{\bot}^{\rm M}(\zeta_l,y)r_{\bot}^{\rm D}(\zeta_l,y)e^{-y}\right]\right\}.
\label{eq1}
\end{eqnarray}
\noindent
Here, the dimensionless Matsubara frequencies are $\zeta_l=\xi_l/\xi_c$,
where $\xi_l=2\pi k_BTl/\hbar$ are the usual Matsubara frequencies,
and $\xi_c=c/(2a)$ is the characteristic frequency, $k_B$ is the
Boltzmann constant, $\delta_{l0}$ is the Kronecker symbol. The reflection 
coefficients for two different polarizations of the electromagnetic
field are expressed in terms of the dielectric permittivities
$\varepsilon^{\rm M,D}(\omega)$ of a metal and a dielectric, respectively,
\begin{eqnarray}
&&
r_{\|}^{\rm M,D}(\zeta_l,y)=\frac{\varepsilon_l^{\rm M,D}y-
\sqrt{y^2+\zeta_l^2(\varepsilon_l^{\rm M,D}-1)}}{\varepsilon_l^{\rm M,D}y+
\sqrt{y^2+\zeta_l^2(\varepsilon_l^{\rm M,D}-1)}},
\nonumber \\
&&r_{\bot}^{\rm M,D}(\zeta_l,y)=\frac{\sqrt{y^2+
\zeta_l^2(\varepsilon_l^{\rm M,D}-1)}
-y}{\sqrt{y^2+\zeta_l^2(\varepsilon_l^{\rm M,D}-1)}+y},
\label{eq2}
\end{eqnarray}
\noindent
where $\varepsilon_l^{\rm M,D}=\varepsilon^{\rm M,D}(i\zeta_l\xi_c)$.

For later use we introduce the so-called effective temperature,
$T_{\rm eff}$, defined through 
$k_BT_{\rm eff}=\hbar\xi_c=\hbar c/(2a)$. In terms of the effective
temperature the nondimensional Matsubara frequencies are expressed as
$\zeta_l=\tau l$, where $\tau$ is a new parameter 
$\tau=2\pi T/T_{\rm eff}$.

For an ideal metal $r_{\|,\bot}^{\rm M}(\zeta_l,y)=1$ and Eq.~(\ref{eq1})
gives us the free energy of the Casimir interaction between an ideal metal
and dielectric
\begin{eqnarray}
&&{\cal F^{\rm ID}}(a,T)=\frac{k_B T}{8\pi a^2}
\sum\limits_{l=0}^{\infty}\left(1-\frac{1}{2}\delta_{l0}\right)
\nonumber \\
&&\phantom{aaa}
\times\int_{\tau l}^{\infty}y\,dy\left\{\ln\left[1-
r_{\|}^{\rm D}(\zeta_l,y)e^{-y}\right]\right.
\nonumber \\
&&\phantom{aaa}
\left.+\ln\left[1-
r_{\bot}^{\rm D}(\zeta_l,y)e^{-y}\right]\right\}.
\label{eq3}
\end{eqnarray}
\noindent
Let us assume that the dielectric is dilute, i.e., having dielectric
permittivity $\varepsilon_l^{\rm D}=1+\eta$, where
$\eta=\mbox{const}\ll 1$.
Expanding the reflection coefficients $r_{\|,\bot}^{\rm D}$ from
Eq.~(\ref{eq2}) in powers of $\eta$, one obtains
\begin{eqnarray}
&&
\ln\left[1-r_{\|}^{\rm D}(\zeta_l,y)e^{-y}\right]=
\eta\frac{e^{-y}}{4}\left(\frac{\tau^2l^2}{y^2}-2\right)
\label{eq4} \\
&&
\phantom{aa}
-\eta^2\frac{e^{-y}}{32y^4}\left[\left(4+e^{-y}\right)\tau^4l^4-
4e^{-y}\tau^2l^2y^2+4\left(e^{-y}-2\right)y^4\right]+
\mbox{O}(\eta^3),
\nonumber \\
&&
\ln\left[1-r_{\bot}^{\rm D}(\zeta_l,y)e^{-y}\right]=
-\eta\frac{e^{-y}}{4}\frac{\tau^2l^2}{y^2}
\nonumber \\
&&
\phantom{aa}
-\eta^2\frac{e^{-y}}{32y^4}\left(e^{-y}-4\right)\tau^4l^4+
\mbox{O}(\eta^3).
\nonumber
\end{eqnarray}

Substituting Eq.~(\ref{eq4}) in Eq.~(\ref{eq3}) and preserving only
two powers in the small parameter $\eta$, we arrive at the following
expression for the Casimir free energy
\begin{eqnarray}
&&{\cal F^{\rm ID}}(a,T)=\frac{\hbar c\tau}{32\pi^2 a^3}
\sum\limits_{l=0}^{\infty}\left(1-\frac{1}{2}\delta_{l0}\right)
\int_{\tau l}^{\infty}y\,dy
\label{eq5} \\
&&\phantom{aa}
\times\left\{-\eta\frac{e^{-y}}{2}+\eta^2\frac{e^{-y}}{16y^4}
\left[4y^4-e^{-y}\left(\tau^4l^4-2\tau^2l^2y^2+2y^4\right)
\right]\right\}.
\nonumber
\end{eqnarray}
\noindent
Upon integrating and summing in Eq.~(\ref{eq5}), we obtain
\begin{eqnarray}
&&{\cal F^{\rm ID}}(a,T)=-\frac{\hbar c}{32\pi^2 a^3}
\left\{
\vphantom{\left[\sum\limits_{l=1}^{\infty}\right]}
\eta\tau\frac{e^{2\tau}+2\tau e^{\tau}-1}{4(e^{\tau}-1)^2}
\right.
\label{eq6} \\
&&
-\eta^2\frac{\tau}{16}
\left[\frac{2\left(e^{2\tau}+2\tau e^{\tau}-1\right)}{(e^{\tau}-1)^2}
-\frac{e^{4\tau}+4\tau e^{2\tau}-1}{4(e^{2\tau}-1)^2}\right.
\nonumber \\
&&
+\tau^2e^{2\tau}
\frac{1-e^{4\tau}+2\tau \left(1+4e^{2\tau}+
e^{4\tau}\right)}{2(e^{2\tau}-1)^4}
\nonumber \\
&&\left.\left.
+2\tau^4\sum\limits_{l=1}^{\infty}l^4\mbox{Ei}(-2\tau l)-
2\tau^2\sum\limits_{l=1}^{\infty}l^2\mbox{Ei}(-2\tau l)
\vphantom{\frac{2\left(e^{2\tau}+2\tau e^{\tau}-1\right]}{(e^{\tau}-1)^2}}
\right]\right\},
\nonumber
\end{eqnarray}
\noindent
where $\mbox{Ei}(z)$ is the integral exponent.

The entropy of the Casimir interaction per unit area of the ideal metal
and dilute dielectric is found as minus the first derivative of 
Eq.~(\ref{eq6}) with respect to the temperature
\begin{eqnarray}
&&S^{\rm ID}(a,T)=
-\frac{\partial{\cal F^{\rm ID}}(a,T)}{\partial T}=
\frac{k_B}{8\pi a^2}\left\{
\frac{\eta}{2}\left[
\vphantom{\sum\limits_{l=1}^{\infty}}
\frac{e^{\tau}+1}{2(e^{\tau}-1)}
\right.\right.
\nonumber \\
&&\left.
-\tau e^{\tau}\frac{1+\tau-(1-\tau)e^{\tau}}{(e^{\tau}-1)^3}
\right]-\frac{\eta^2}{16}\left[\frac{7e^{\tau}+9}{4(e^{\tau}-1)}
\right.
\label{eq7} \\
&&
-4\tau e^{\tau}\frac{1+\tau-(1-\tau)e^{\tau}}{(e^{\tau}-1)^3}
-\frac{2\tau e^{2\tau}+e^{2\tau}-1}{2(e^{2\tau}-1)^2}
\nonumber \\
&&
+\tau^2 e^{2\tau}
\frac{3(1-e^{4\tau})+10\tau(1+4e^{2\tau}+e^{4\tau})}{2(e^{2\tau}-1)^4}
\nonumber \\
&&
\left.\left.
-\sum\limits_{l=1}^{\infty}\left(6\tau^2l^2-10\tau^4l^4\right)
\mbox{Ei}(-2\tau l)
\vphantom{\frac{3(e^{4\tau})+10\tau(4e^{2\tau}+e^{4\tau}}{2(e^{2\tau})^4}}
\right]\right\}.
\nonumber
\end{eqnarray}
\noindent
Eq.~(\ref{eq6}) permits to find an analytic expression for the
Casimir pressure between the ideal metal and dilute dielectric
\begin{eqnarray}
&&P^{\rm ID}(a,T)=
-\frac{\partial{\cal F^{\rm ID}}(a,T)}{\partial a}=
-\frac{\hbar c\tau}{32\pi^2 a^4}\left\{
\vphantom{\left[\sum\limits_{l=1}^{\infty}\right]}
\frac{\eta}{2}\left[
\vphantom{\sum\limits{l=1}^{\infty}}
\frac{e^{2\tau}+2\tau e^{\tau}-1}{(e^{\tau}-1)^2}
\right.\right.
\nonumber \\
&&\left.
+\frac{\tau^2 e^{\tau}(e^{\tau}+1)}{(e^{\tau}-1)^3}
\vphantom{\sum\limits{l=1}^{\infty}}
\right]-\frac{\eta^2}{4}\left[
\vphantom{\sum\limits{l=1}^{\infty}}
\frac{e^{2\tau}+2\tau e^{\tau}-1}{(e^{\tau}-1)^2}
\right.
\label{eq8} \\
&&
+\frac{\tau^2 e^{\tau}(e^{\tau}+1)}{(e^{\tau}-1)^3}
-\frac{e^{4\tau}+4\tau e^{2\tau}-1}{8(e^{2\tau}-1)^2}
\nonumber \\
&&
-\frac{\tau^3e^{2\tau}\left(e^{4\tau}+4e^{2\tau}+
1\right)}{2(e^{2\tau}-1)^4}
\left.\left.
-\tau^4\sum\limits_{l=1}^{\infty}l^4
\mbox{Ei}(-2\tau l)
\vphantom{\frac{3(e^{4\tau})+10\tau(4e^{2\tau}+e^{4\tau}}{2(e^{2\tau})^4}}
\right]\right\}.
\nonumber
\end{eqnarray}

Note that Eqs.~(\ref{eq6})--(\ref{eq8}) are perturbative in the small parameter
$\eta$ but for each perturbation order the thermal effects are taken into
account exactly. This gives the possibility to find the high-temperature
behavior ($\tau\gg 1$) of the Casimir free energy, entropy and pressure.
Thus, from Eq.~(\ref{eq6}) in the limit $\tau\gg 1$ ($T\gg T_{\rm eff}$)
for the free energy it follows
\begin{equation}
{\cal F^{\rm ID}}(a,T)=-\frac{k_B T}{32\pi a^2}\eta
\left(1-\frac{7}{16}\eta\right).
\label{eq9}
\end{equation}
\noindent
In a similar manner from  Eqs.~(\ref{eq7}) and (\ref{eq8}) one finds
the behavior of the Casimir entropy and pressure at high temperature
\begin{eqnarray}
&&
{S^{\rm ID}}(a,T)=\frac{k_B}{32\pi a^2}\eta
\left(1-\frac{7}{16}\eta\right),
\nonumber \\
&&
{P^{\rm ID}}(a,T)=-\frac{k_BT}{16\pi a^3}\eta
\left(1-\frac{7}{16}\eta\right).
\label{eq10}
\end{eqnarray}
\noindent
Needless to say that entropy and pressure from Eq.~(\ref{eq10})
can be obtained also as minus the derivatives of Eq.~(\ref{eq9}) with
respect to temperature and separation distance, respectively.

Now let us consider the low temperature behavior of the Casimir free 
energy, entropy and pressure.
For this purpose the direct application of Eqs.~(\ref{eq6})--(\ref{eq8})
is rather cumbersome. The same results can be obtained more simply 
by using the Abel-Plana formula (see, e.g., Refs.~\cite{3,20})
\begin{eqnarray}
&&
\sum\limits_{l=0}^{\infty}\left(1-\frac{1}{2}\delta_{l0}\right)
F(l)=
\int_{0}^{\infty}F(t)dt
\label{eq11} \\
&&\phantom{aa}
+i\int_{0}^{\infty}
\frac{dt}{e^{2\pi t}-1}\left[F(it)-F(-it)\right],
\nonumber
\end{eqnarray}
\noindent
where $F(z)$ is an analytic function on the right half-plane.
Before the application of Eq.~(\ref{eq11}), we return to Eq.~(\ref{eq5})
and perform the integration with respect to $y$ (but not the summation
over $l$). The result is
\begin{equation}
{\cal F^{\rm ID}}(a,T)=-\frac{\hbar c\tau}{32\pi^2 a^3}
\sum\limits_{l=0}^{\infty}\left(1-\frac{1}{2}\delta_{l0}\right)
F(l),
\label{eq12}
\end{equation}
\noindent
where
\begin{eqnarray}
&&
F(l)=\frac{\eta}{2}(1+\tau l)e^{-\tau l}-
\frac{\eta^2}{16}\left[
\vphantom{\frac{1}{2}}
4(1+\tau l)e^{-\tau l}\right.
\label{eq13} \\
&&
\left.
-\frac{1}{2}\left(1+2\tau l+\tau^2l^2-2\tau^3l^3\right)e^{-2\tau l}
-2\tau^2l^2\left(1-\tau^2l^2\right)\mbox{Ei}(-2\tau l)\right].
\nonumber
\end{eqnarray}

The sum in Eq.~(\ref{eq12}) can be calculated using the Abel-Plana
formula (\ref{eq11}). In doing so, we note that the first term on
the right-hand side of Eq.~(\ref{eq11}), where $F(t)$ is given by
Eq.~(\ref{eq13}) with a substitution of $t$ for $l$, 
is proportional to the Casimir energy at zero temperature,
whereas the second gives the thermal correction. Direct integration 
results in
\begin{equation}
\int_{0}^{\infty}F(t)dt=\frac{\eta}{\tau}
\left(1-\frac{457}{960}\eta\right).
\label{eq14}
\end{equation}
\noindent
The second integral on the right-hand side of Eq.~(\ref{eq11})
is calculated as a perturbative expansion in powers of a small
parameter $\tau$ with the aim to obtain the low temperature
asymptotics. Preserving all terms up to $\tau^3$ inclusive, we
obtain from Eq.~(\ref{eq13})
\begin{eqnarray}
&&
F(it)-F(-it)=-\frac{i\eta}{3}\left[\tau^3t^3+\mbox{O}(\tau^5)
\right]
\label{eq15} \\
&&
+\frac{i\eta^2}{8}\left[-\pi\tau^2t^2+ 6\tau^3t^3+\mbox{O}(\tau^4)
\right].
\nonumber
\end{eqnarray}
\noindent
Using this equation, the integral can be calculated
\begin{eqnarray}
&&
i\int_{0}^{\infty}
\frac{dt}{e^{2\pi t}-1}\left[F(it)-F(-it)\right]
\label{eq16} \\
&&
\phantom{aa}
=\frac{\eta\tau^3}{720}-\frac{\eta^2\tau^2}{32}\left[
\frac{\tau}{10}-\frac{\zeta(3)}{\pi^2}\right],
\nonumber
\end{eqnarray}
\noindent
where $\zeta(z)$ is the Riemann zeta function.

Substituting Eqs.~(\ref{eq11}), (\ref{eq14}), (\ref{eq16}) in
Eq.~(\ref{eq12}), the low temperature behavior of the free energy
is obtained
\begin{eqnarray}
&&
{\cal F^{\rm ID}}(a,T)=-\frac{\hbar c}{32\pi^2 a^3}\eta
\left\{
\vphantom{\left[\frac{\tau^4}{10}\right]}
1+\frac{\tau^4}{720}\right.
\label{eq17} \\
&&\left.
\phantom{aa}-\frac{\eta}{32}\left[
\frac{457}{30}-\frac{\zeta(3)\tau^3}{\pi^2}+\frac{\tau^4}{10}\right]
\right\}
\nonumber
\end{eqnarray}
\noindent
[here the terms of order $\mbox{O}(\eta\tau^6)$ and 
$\mbox{O}(\eta^2\tau^5)$ are omitted].

The low temperature asymptotics of the Casimir entropy and pressure 
can be obtained by the calculation of minus the derivatives of
Eq.~(\ref{eq12}) (with respect to temperature and separation,
respectively) and by subsequent application of the Abel-Plana
formula with the resulting functions $F(l)$. The same asymptotic
expressions are obtained also by direct differentiation of
Eq.~(\ref{eq17}). They are as follows:
\begin{eqnarray}
&&
{S^{\rm ID}}(a,T)=\frac{k_B}{32\pi a^2}\eta\tau^2
\left\{
\vphantom{\left[\frac{\tau^4}{10}\right]}
\frac{\tau}{45}+\frac{\eta}{4}\left[
\frac{3\zeta(3)}{2\pi^2}-\frac{\tau}{5}\right]
\right\},
\label{eq18} \\
&&
{P^{\rm ID}}(a,T)=-\frac{\hbar c}{32\pi^2 a^4}\eta
\left[
\vphantom{\left[\frac{\tau^4}{10}\right]}
3-\frac{\tau^4}{720}-\frac{\eta}{320}\left(
457-\tau^4\right)
\right].
\nonumber
\end{eqnarray}
\noindent

It is significant that $S^{\rm ID}\to 0$ when $T\to 0$ (recall that
$\tau\sim T$ by definition), so that the Nernst heat theorem is
satisfied. For dilute dielectrics from Eqs.~(\ref{eq7}), (\ref{eq18})
it follows also that $S^{\rm ID}\geq 0$.
At the same time, there are temperature intervals 
where the Casimir pressure is nonmonotonous.
To demonstrate this, in Fig.~1 the relative thermal correction to
the Casimir pressure, defined as 

\begin{equation}
\delta_TP=
\frac{{P}(a,T)-
{P}(a,0)}{{P}(a,0)},
\label{eq19}
\end{equation}
\noindent
is plotted as a function of temperature for $\eta=0.001$ (solid
line) and $\eta=0.1$ (dashed line) at a separation distance
$a=2\,\mu$m [here we use Eq.~(\ref{eq8}) and $P(a,T)=P^{\rm ID}(a,T)$]. 
As is seen from Fig.~1, the region where the thermal correction is 
negative is in accordance with the results of Ref.~\cite{21}
for the Casimir-Polder energy and force between an atom and a
metal wall.

From Fig.~1 it is seen that for $\eta=0.1$ it holds 
$\delta_TP^{\rm ID}<0$ at 
temperatures $T\leq 343\,$K. In this case he relative thermal correction 
has the minimum value $\delta_TP^{\rm ID}\approx -0.007$ at
$T=270\,$K. 
For comparison, at $T=400\,$K the thermal correction
has the positive value $\delta_TP^{\rm ID}\approx 0.018$.
Note that there is some analogy between the behavior of the Casimir 
pressure and thermal correction as a function of temperature,
obtained here on the one hand, and discussed in Refs.~\cite{24,25,26,27} 
on the other hand in the case of real metals described by the 
Drude dielectric function. We will return to this analogy in more 
detail in Sec.~V after the extension of the above results 
for more realistic plate materials which will be obtained in the 
following sections.

\section{Thermal corrections to the Casimir interaction between
an ideal metal and dielectric with frequency-independent
dielectric permittivity}

In this section we consider one plate made of dielectric with arbitrary
constant $\varepsilon^{\rm D}$ (not dilute) and the other plate made of ideal
metal as before. The application regions of the asymptotic expressions
at low and high temperatures, obtained in Sec.~II, will be our
initial concern. To gain an impression of how
accurate the asymptotic of Eq.~(\ref{eq17}) is, let us calculate the
relative thermal correction to the Casimir energy,
\begin{equation}
\delta_T{\cal F}=
\frac{{\cal F}(a,T)-
{\cal F}(a,0)}{{\cal F}(a,0)},
\label{eq19a}
\end{equation}
\noindent
as a function of temperature
both numerically [by the direct use of the Lifshitz formula (\ref{eq3})]
and analytically [using Eq.~(\ref{eq17})].
In Fig.~2 the numerical results are presented by the solid line and 
the analytic ones by the dashed line for a dielectric with 
$\varepsilon_0^{\rm D}=1.1$ ($\eta=0.1$) at a separation $a=1\,\mu$m.
As is seen from this figure, at $T<220\,$K there is practically exact
coincidence of numerical and analytic results. What this means is that
at $\tau\leq 1.2$ (the effective temperature here is equal to
$T_{\rm eff}=1145\,$K) one can use the asymptotic expression of 
Eq.~(\ref{eq17}) 
in order to get accurate results.

The asymptotic regime of high temperatures is given by Eq.~(\ref{eq9}).
This regime is achieved at $T=1500\,$K where the values
of $\delta_T{\cal F}^{\rm ID}$, 
calculated by Eqs.~(\ref{eq3}) and (\ref{eq9}),
differ for only 0.2\%. For higher temperatures ($\tau\geq 8.2$)
Eq.~(\ref{eq9}) is applicable for accurate calculations.

We now turn to the dielectrics with larger $\varepsilon_0^{\rm D}$.
For the sake of convenience in the comparison of experiment and theory,
we will present the results of the calculation as a function of separation
rather than temperature. The point is that in all the previous experiments
of Refs.~\cite{5,6,7,8,9,10,11,12,13,14} the temperature was constant
($T=300\,$K) and the separation distance was variable (there is only one
proposed but not yet completed experiment \cite{43,43a} exploiting the effect
of two different temperatures). Theoretically, the two asymptotic regimes
of low ($T\ll T_{\rm eff}$) and high ($T\gg T_{\rm eff}$) temperatures
are equivalent to the limits of small and large separations, respectively.
This becomes evident if it is recalled that $T_{\rm eff}\sim a^{-1}$.
We consider the separation region $100\,\mbox{nm}\leq a\leq 1400\,$nm
where the nonmonotonous behavior of the thermal correction to the Casimir 
energy is observed for dielectrics with larger $\varepsilon$ (not dilute). 
In fact, even in that separation region effects 
which are due to the nonideality of a metal and the 
absorption bands 
in a dielectric (see, e.g., Refs.~\cite{44,45}) become important.
At shorter separations ($a< 100\,$nm) these effects lead to the result 
that the model of an ideal metal and a dielectric with constant dielectric
permittivity becomes inapplicable (see Sec.~IV where the case of real metal 
and dielectric is considered).

In Fig.~3a we present the values of the 
relative thermal correction (\ref{eq19a})
to the Casimir energy at $T=300\,$K as a function of separation
computed by the use of Lifshitz formula (\ref{eq3}) for ideal metal
and dielectrics with different dielectric permittivities
$\varepsilon_0^{\rm D}=3,\,6,\,7,\,10$ (lines 1, 2, 3, and 4, respectively).
As is seen from Fig.~3a, lines 3 and 4 demonstrate the nonmonotonous
behavior of the thermal correction to the free energy (recall that
in the case of an ideal metal and dilute dielectric only the Casimir
pressure is a nonmonotonous function of the temperature, see
Fig.~1). Curiously, line 3 possesses both a minimum and 
a maximum values within the separation interval under consideration.

For line 4 ($\varepsilon_0^{\rm D}=10$) the zero value of the thermal
correction is achieved at $a_1=0.2\,\mu$m and at $a_2=1.25\,\mu$m.
Within the separation region $a_1<a<a_2$ the relative thermal correction is 
negative. Using the respective effective temperatures 
($T_{\rm eff}^{(1)}=5724\,$K at $a=a_1$ and  $T_{\rm eff}^{(2)}=916\,$K at 
$a=a_2$) we find that in terms of the dimensionless parameter $\tau$,
introduced in Sec.~II, the thermal correction is negative within
the interval $0.33\leq\tau\leq 2.06$. The minimum on line 4 is
achieved at $a=0.9\,\mu$m ($\tau=1.48$). What this means is that the
entropy of a fluctuating field takes negative values within the interval 
$0.33\leq\tau\leq 1.48$ (in the case of a plate made of dilute
dielectric the entropy is positive). 
In terms of usual temperatures for the plates
at a separation $a=0.2\,\mu$m, the Casimir entropy is negative within
the interval $300\,\mbox{K}\leq T\leq 1350\,$K (if, as assumed in this
section, $\varepsilon^{\rm D}$ is independent of the temperature).
Note that for the dielectric with $\varepsilon_0^{\rm D}=10$ the thermal
correction is almost 0.5\% at $a=0.9\,\mu$m, i.e., much greater than in 
the case of a dilute dielectric. It remains, however, still too small
to be observed experimentally.

To illustrate the behavior of the Casimir entropy as a function of
temperature in the configuration of a plate made of ideal metal and
another plate made of dielectric with a frequency-independent
dielectric permittivity, we plot it in Fig.~3b for
$\varepsilon_0^{\rm D}=7$ and separation distance $a=600\,$nm. As is seen
from Fig.~3b, the entropy is negative within the interval from
137\,K to 311\,K (i.e., from 0.45 to 1.02 in terms of $\tau$).
The minimum value of the entropy equal to
$-14\,\mbox{KeV\,m}^{-2}\,\mbox{K}^{-1}$ is
achieved at a temperature $T=238\,$K ($\tau=0.78$). When this result
is compared with the above case of $\varepsilon_0^{\rm D}=10$, it is
apparent that the region, where the entropy is negative, is
narrowed with the decrease of the dielectric permittivity.
Thus, for $\varepsilon_0^{\rm D}<6$ the Casimir entropy is already 
positive at any temperature.

The negativeness of the Casimir entropy within some temperature
(separation) intervals should not become of concern.
It is self-evident that the entropy of a closed
system, which includes the space occupied by the dielectric plate, is
positive (the second plate is made of an ideal metal with the
Dirichlet boundary condition on it; for this reason it does not
contribute to the entropy of the system under consideration).
It should be noted also that the Nernst heat theorem is satisfied 
perfectly good. To proof this fact analytically, we apply the
Abel-Plana formula (\ref{eq11}) in Eq.~(\ref{eq3}) where the role
of $F(l)$ is played by the integral in the right-hand side of
Eq.~(\ref{eq3}). Preserving only the lowest expansion order in
the parameter $\tau l$, we arrive at
\begin{eqnarray}
&&
F(l)=\tau^2l^2\frac{\varepsilon_0^{\rm D}-1}{\varepsilon_0^{\rm D}+1}
\int_{\tau l}^{\infty}dy
\label{eq19b} \\
&&\phantom{aa}
\times
\left[
\frac{\varepsilon_0^{\rm D}}{\left(\varepsilon_0^{\rm D}+1\right)-
\left(\varepsilon_0^{\rm D}-1\right)e^{-y}}-
\frac{\varepsilon_0^{\rm D}+1}{4}\right]
\frac{e^{-y}}{y}+O\left(\tau^3l^3\right).
\nonumber
\end{eqnarray}
\noindent
From this after integration it follows
\begin{equation}
F(it)-F(-it)=i\pi\tau^2t^2
\frac{\left(\varepsilon_0^{\rm D}-1\right)^2}{4\left(
\varepsilon_0^{\rm D}+1\right)}.
\label{eq19c}
\end{equation}
\noindent
Performing the integration with respect to $t$ in Eq.~(\ref{eq11}),
we obtain from Eq.~(\ref{eq3}) the asymptotic behavior of the
Casimir free energy
\begin{equation}
{\cal F}^{\rm ID}(a,T)={\cal F}^{\rm ID}(a,T=0)-
\frac{\hbar c\zeta(3)\tau^3}{512\pi^4a^3}\,
\frac{\left(\varepsilon_0^{\rm D}-1\right)^2}{\varepsilon_0^{\rm D}+1}.
\label{eq19d}
\end{equation}
\noindent
Note that for dilute dielectric this coincides with the term of
Eq.~(\ref{eq17}) of order $\tau^3$ as it should be.

From Eq.~(\ref{eq19d}) we finally arrive to the low-temperature
asymptotics of the entropy
\begin{equation}
{S}^{\rm ID}(a,T)=
\frac{3k_B\zeta(3)\tau^2}{128\pi^3a^2}\,
\frac{\left(\varepsilon_0^{\rm D}-1\right)^2}{\varepsilon_0^{\rm D}+1},
\label{eq19e}
\end{equation}
\noindent
which is also in perfect agreement with the second-order term
in Eq.~(\ref{eq18}) obtained for dilute dielectrics.
From Eq.~(\ref{eq19e}) it follows that the Casimir entropy
goes to zero as the second power of the temperature, i.e.,
the Nernst heat theorem is satisfied. The comparison with the
numerical computations shows that the asymptotic expressions
(\ref{eq19d}), (\ref{eq19e}) work good for all $\tau\leq 0.1$.

\section{Thermal corrections to the Casimir interaction between
real metal and dielectric}

In this section we consider one of the plates made of real metal (Au) 
and the other plate made of a real dielectric (Si or 
$\alpha\mbox{-Al}_2\mbox{O}_3$). Both these chosen dielectrics possess
relatively large values of the static dielectric permittivity and
quite different behavior of $\varepsilon^{\rm D}(i\xi)$ around the
characteristic frequencies for the separations under consideration.
The Casimir free energy is found by the use of the complete
Lifshitz formula (\ref{eq1}) describing the case of two parallel
plates made of real materials. The dielectric permittivity of real
materials along the imaginary frequency axis can be obtained through
the dispersion relation
\begin{equation}
\varepsilon^{\rm M,D}(i\xi)=1+\frac{2}{\pi}
\int_{0}^{\infty}d\omega
\frac{\omega\;\mbox{Im}\,\varepsilon^{\rm M,D}(\omega)}{\omega^2+\xi^2}.
\label{eq20}
\end{equation}
\noindent
Here the imaginary part of the dielectric permittivity is calculated
as $2n_1n_2$ where $n_1$ and $n_2$ are the real and imaginary parts
of the complex refraction index tabulated, e.g., in Ref.~\cite{46}.
For Au the available tabulated data are extended for lower frequencies 
using the usual procedure (see, e.g., Refs.~\cite{44,45}).
The resulting behavior of $\varepsilon^{\rm Au}$ as a function of
$\xi$ can be found in Refs.~\cite{20,23,44,45}.
The same procedure, applied in the case of Si (here the tabulated data
for lower frequencies, than for Au, are available so that no
additional extension of data is needed), leads to the results shown
by line 1 in Fig.~4.

There are also good analytic formulas describing the behavior of the
dielectric permittivities of different materials along the
imaginary frequency axis. As an example, the dielectric permittivity
of $\alpha\mbox{-Al}_2\mbox{O}_3$ is well described \cite{41} in the
Ninham-Parsegian representation \cite{1} 
\begin{equation}
\varepsilon^{\rm D}(i\xi)=1+\frac{C_{\rm IR}}{1+
\frac{\xi^2}{\omega_{\rm IR}^2}}+\frac{C_{\rm UV}}{1+
\frac{\xi^2}{\omega_{\rm UV}^2}},
\label{eq21}
\end{equation}
\noindent
where $\omega_{\rm IR}=1\times 10^{14}\,$rad/s,
$\omega_{\rm UV}=2\times 10^{16}\,$rad/s are the characteristic
absorption frequencies, and $C_{\rm IR}=7.03$, $C_{\rm UV}=2.072$
are the corresponding absorption strengths in the infrared and
ultraviolet ranges, respectively. The dielectric permittivity
of $\alpha\mbox{-Al}_2\mbox{O}_3$ as a function of $\xi$ is plotted
in Fig.~4, line 2.
As is seen from Fig.~4, the dielectric permittivities of Si and
$\alpha\mbox{-Al}_2\mbox{O}_3$ are qualitatively different in the
region of characteristic frequencies $\xi_c\sim 10^{15}\,$rad/s.
In fact, for $\alpha\mbox{-Al}_2\mbox{O}_3$ in the region around
$\xi_c$ the values of $\varepsilon^{\rm D}$ correspond to the second step
of line 2 and are several times less than $\varepsilon_0^{\rm D}=10.1$,
whereas for Si the static value of the dielectric permittivity
$\varepsilon_0^{\rm D}=11.66$ is preserved up to the region of 
characteristic frequencies.

It should be stressed that the computations below are unaffected by 
the controvercies concerning the contribution of the zero-frequency
term of the Lifshitz formula in the case of real metals (this
contribution is different in the approaches using the Drude dielectric
function and the Leontovich surface impedance, see Introduction).
The reason is that in our case only one plate is made of a real
metal, whereas the other one is made of a dielectric. According to
Eq.~(\ref{eq2}), for dielectrics with finite
$\varepsilon^{\rm D}(0)=\varepsilon_0^{\rm D}$ it follows
\begin{equation}
r_{\|}^{\rm D}(0,y)=
\frac{\varepsilon_0^{\rm D}-1}{\varepsilon_0^{\rm D}+1},
\qquad
r_{\bot}^{\rm D}(0,y)=0.
\label{eq22}
\end{equation}
\noindent
As a result, if one plate is made of a dielectric, the transverse
electric mode at zero frequency does not contribute to the Casimir
free energy regardless of the approach used to describe the metal
of the other plate [i.e., regardless of the value of the transverse
reflection coefficient $r_{\bot}^{\rm M}(0,y)$].

A further distinctive feature of the different approaches to the thermal 
Casimir force in the case of real metal, which might play a part in
determining the contribution to the free energy at nonzero Matsubara
frequencies, is the form of reflection coefficients. In the framework
of the impedance approach \cite{28,33}, in place of the usual
reflection coefficients (\ref{eq2}), expressed in terms of the 
dielectric permittivity, one should use the coefficients, expressed
in terms of the Leontovich surface impedance. This, however, does not 
present a problem because, as was demonstrated in Ref.~\cite{14}, at all
nonzero Matsubara frequencies the contributions from both types of
the reflection coefficients are practically the same.

In Fig.~5 we present the results of the calculation for the relative
thermal correction to the Casimir energy between Au and Si plates 
obtained by Eqs.~(\ref{eq1}) and (\ref{eq2}) using the procedure
described above  at $T=300\,$K (solid line). For comparison, in the
same figure the dashed line shows the results obtained using the approach
of Sec.~III (i.e., for ideal metal and the dielectric with a 
frequency-independent dielectric permittivity 
$\varepsilon_0^{\rm D}=11.66$ equal to the
static permittivity of Si). As is seen from Fig.~5 (solid line),
there is a wide separation interval 
$0.2\,\mu\mbox{m}\leq a\leq 1.3\,\mu$m where the relative thermal correction
to the Casimir energy in the case of real materials is negative
(in terms of the dimensionless variable 
this holds for $0.33\leq\tau\leq 2.14$).
The minimum value of the thermal correction $\delta_T{\cal F}=-0.006$ is
achieved at $a=0.95\,\mu$m ($\tau=1.56$). What this means is that
the Casimir entropy in the case of real materials is negative
within the separation region $0.2\,\mu\mbox{m}\leq a\leq 0.95\,\mu$m
(or, in terms of $\tau$, for $0.33\leq\tau\leq 1.56$).
The comparison with the dashed line shows that for Si the simple
model, used in Sec.~III, leads to the same qualitative results with
only minor differences in the minimum values of $\delta_T{\cal F}$ and the
width of the intervals where the thermal correction and the Casimir
entropy are negative.

We now turn to the Casimir interaction of Au plate with a plate made 
of $\alpha\mbox{-Al}_2\mbox{O}_3$. As was discussed above, the behavior 
of the dielectric permittivity of $\alpha\mbox{-Al}_2\mbox{O}_3$ along the 
imaginary frequency axis is different from that of Si. The results of the 
calculation for the relative thermal correction 
to the Casimir energy as a function of separation 
at $T=300\,$K, obtained by Eqs.~(\ref{eq1}), (\ref{eq2}) and (\ref{eq21}),
are shown in Fig.~6 by the solid line. The dashed line is calculated
for an ideal metal and a dielectric with the frequency-independent
dielectric permittivity $\varepsilon_0^{\rm D}=10.1$ equal to the static
dielectric permittivity of $\alpha\mbox{-Al}_2\mbox{O}_3$. As is seen
from Fig.~6, in this case the solid line presents the monotonously 
increasing positive function of the separation distance. The respective
Casimir entropy is also nonnegative within the separation region
reflected in the figure. The application of the simplified
model of Sec.~III to $\alpha\mbox{-Al}_2\mbox{O}_3$ leads to
qualitatively different results shown by the dashed line in Fig.~6.
This line demonstrates the negative thermal correction within the
separation region from  $a_1=0.25\,\mu$m to  $a_2=1.27\,\mu$m and the
negative Casimir entropy within the separations from 0.25 to
 $0.9\,\mu$m. Thus, the use of realistic data for the 
dielectric permittivities of the plates is essential for the final results.

\section{Conclusions and discussion}

In the foregoing we have investigated the thermal corrections to the
Casimir interaction between metallic and dielectric plates. This was
done both analytically (using the idealized model of an ideal metal
and a dilute dielectric) and numerically (for the ideal metal and
dielectric with a frequency independent dielectric permittivity, and
for real metal and two different dielectrics with dissimilar 
behavior of their dielectric
permittivities along the imaginary 
frequency axis). The main conclusion obtained
above is that the pressure and the
free energy of the Casimir interaction between metal
and dielectric plates may be a nonmonotonous functions of the temperature
within some definite regions. This leads to the possibility of
negative relative thermal corrections and negative values of entropy of
the fluctuating field (the latter holds only for dielectrics with 
sufficiently large dielectric permittivity).
Using the proximity force theorem, one can
conclude that the relative thermal correction to the Casimir force
between a plane metal plate and a spherical dielectric
lens (the configuration used
in many experiments) also can be negative.

The physical interpretation of the obtained results is based on the
fact that both the free energy and entropy of the closed system under
consideration consist of contributions from the plates 
and from their interaction (in the previous sections the latter 
were denoted as ${\cal F}^{\rm ID}$,
$S^{\rm ID}$ or as ${\cal F}$, $S$ for real materials). 
The above conclusions about the possibility of a
nonmonotonous behavior of the free energy and of the negativeness
of the entropy are relevant not for the closed system but due to the
interaction between its parts. In the case of two plates made of an
ideal metal with the Dirichlet boundary conditions on their surface,
there is no penetration of the fluctuating field inside the plates.
In this case the characteristics of the closed system coincide with
those obtained for the interaction between the plates. As a result,
for ideal metals
the free energy of a fluctuating field is a monotonous function
and the entropy is positive. For two dielectric plates \cite{26,47}
or for one dielectric and one metal plate this is not necessarily so.

It is important to keep in mind that only the interaction parts of the
free energy and entropy depend on a separation distance:
${\cal F}={\cal F}(a,T)$, $S=S(a,T)$.
This leads to two conclusions of considerable significance.
The first is that the thermal correction to the Casimir force
(which is minus the derivative of the free energy with respect to
the separation) can be negative. The second is that the Nernst heat
theorem must be valid separately for the contribution
to the entropy from the interaction between the plates, 
so that $S(a,0)=0$, and for the entropy
of the plates. If this were not the case,
i.e., if the equation $S(a,0)=f(a)\neq 0$ were valid (like
in Refs.~\cite{24,25,26,27} for perfect crystal with no impurities), 
then the Nernst heat theorem for the closed system would be violated 
as the entropies of the plates do not depend on $a$. Both these
conclusions were illustrated above by the example of the
Casimir interaction between metal and dielectric.

To conclude, the nonmonotonous dependence of the Casimir free energy 
on temperature and the negative values of the relative thermal correction
(as, for instance, was predicted for real metals in the approach
of Refs.~\cite{24,25,26,27} and for Si in Ref.~\cite{47}) are not
in themselves excluded thermodynamically. Such behavior for real
metals is, however, unlikely because there is only a small 
penetration of the electromagnetic fluctuations at the characteristic
frequencies in the interior of a metal (recall that in the approach
of Refs.~\cite{28,33}, where this property of real metals is taken into
account, the free energy is monotonous, and the relative thermal 
correction is positive). The decisive theoretical argument to give
preference to any approach is, thus, the fulfillment of the Nernst
heat theorem for the Casimir entropy of a fluctuating field for both
perfect crystals and crystals with impurities. In particular, if
one interacting body is made of a metal and the other of a dielectric, 
the entropy of the fluctuating field vanishes when the temperature goes 
to zero.

\section*{Acknowledgments}

G.L.K. and V.M.M. are grateful to the Center of Theoretical Studies and
the Institute for Theoretical
Physics, Leipzig University for their kind hospitality. 
This work was supported by Deutsche Forschungsgemeinschaft grant 
436\,RUS\,113/789/0-1. G.L.K. and V.M.M. were also partially
supported by Finep (Brazil).


\begin{figure*}
\vspace*{-8cm}
\includegraphics{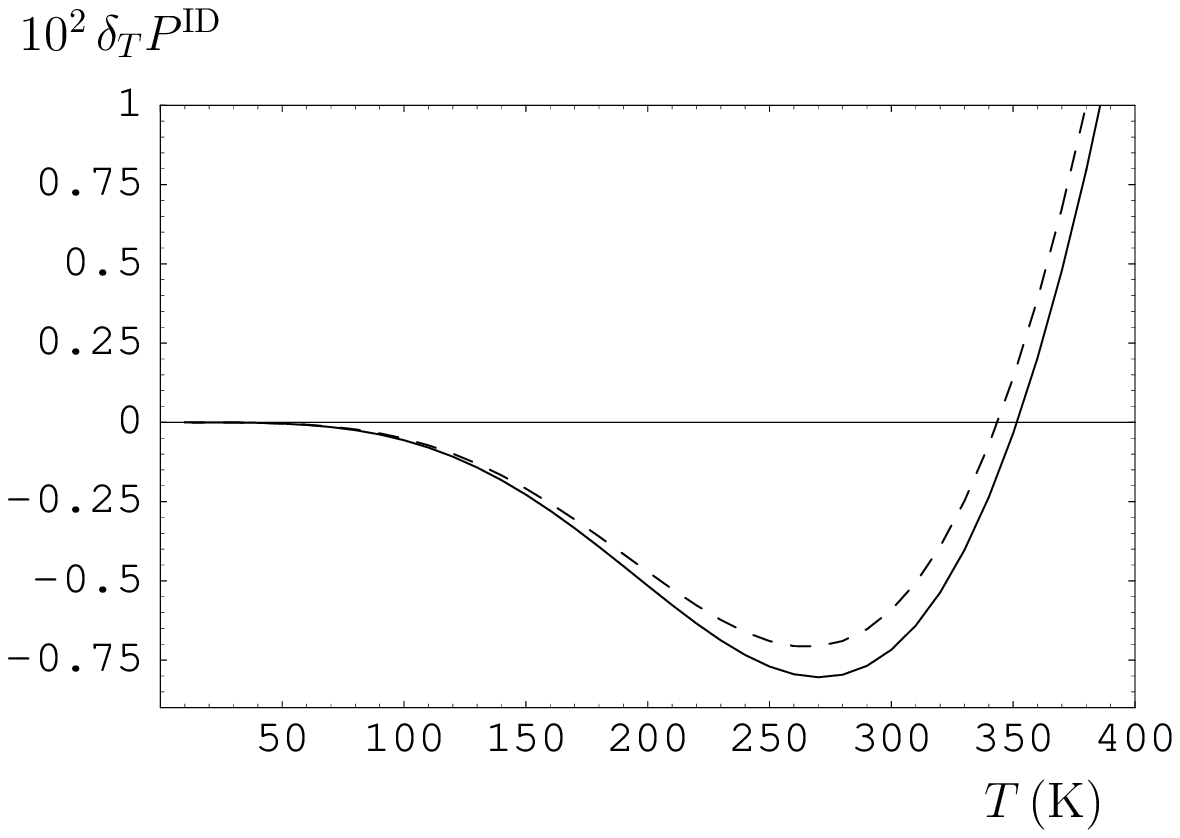}
\vspace*{-9cm}
\caption{
Relative thermal correction to the
Casimir pressure as a function of temperature at a separation
$a=2\,\mu$m for two plates, one  made of an ideal 
metal and the other of dilute dielectric with $\varepsilon_0^{\rm D}=1.001$
(solid line) and $\varepsilon_0^{\rm D}=1.1$ (dashed line).
}
\end{figure*}
\begin{figure*}
\vspace*{-8cm}
\includegraphics{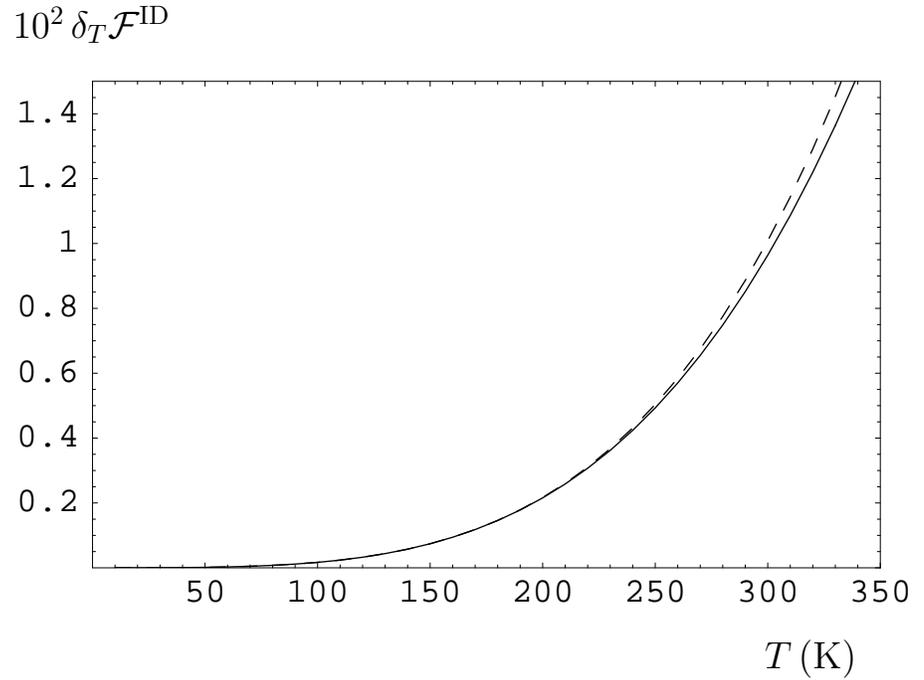}
\vspace*{-9cm}
\caption{
Relative thermal correction to the Casimir energy at a separation
$a=1\,\mu$m as a function of temperature for two plates,
one  made of an ideal metal and the other of
dilute dielectric with $\varepsilon_0^{\rm D}=1.1$, computed
numerically
(solid line) and by the asymptotics of low temperatures (dashed line).
}
\end{figure*}

\begin{figure*}
\vspace*{-3.5cm}
\includegraphics{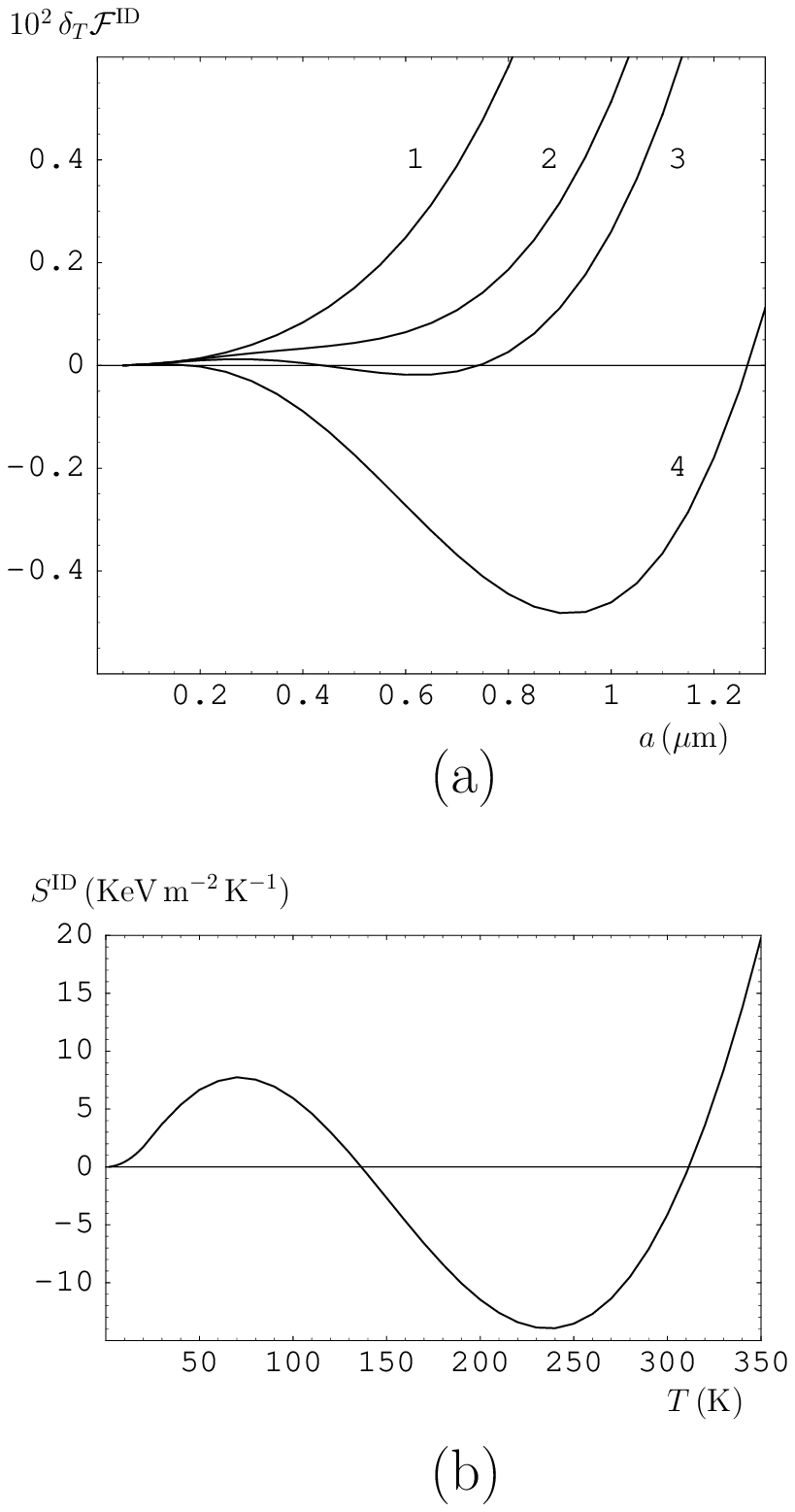}
\vspace*{-12.5cm}
\caption{
(a) Relative thermal correction to the Casimir energy at
$T=300\,$K as a function of separation for two plates,
one made of an ideal metal and the other of 
dielectrics with $\varepsilon_0^{\rm D}=3$, 6, 7, and 10
(lines 1, 2, 3, and 4, respectively);
(b) Casimir entropy as a function of temperature for the
dielectric plate with  $\varepsilon_0^{\rm D}=7$ at
a separation $a=600\,$nm from a metal plate.
}
\end{figure*}
\begin{figure*}
\vspace*{-8cm}
\includegraphics{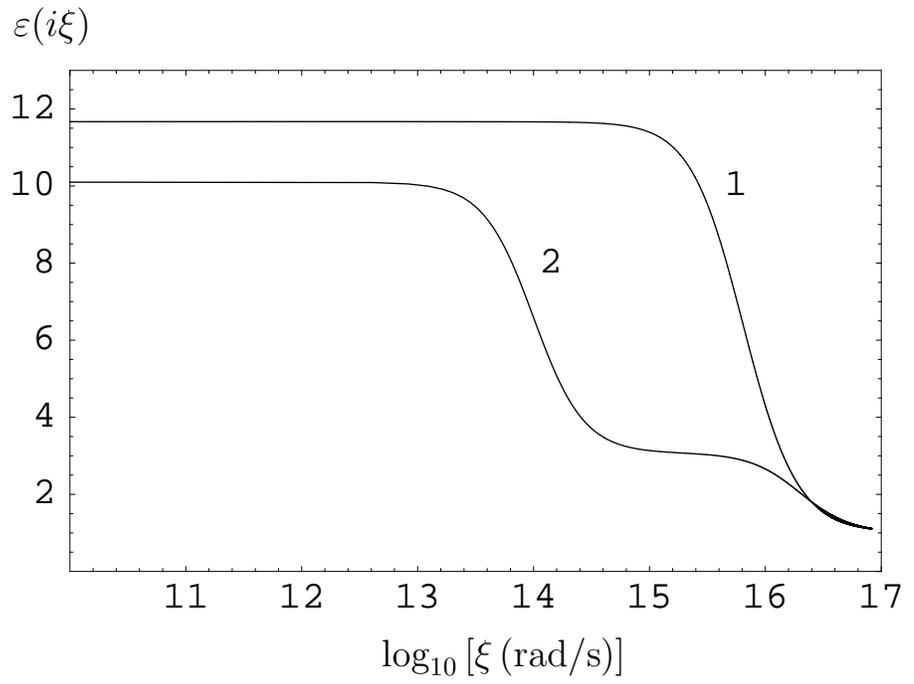}
\vspace*{-9cm}
\caption{
Dielectric permittivity of Si (line 1) and
$\alpha\mbox{-Al}_2\mbox{O}_3$ (line 2) along
the imaginary frequency axis as a function of the logarithm
of frequency. 
}
\end{figure*}
\begin{figure*}
\vspace*{-8cm}
\includegraphics{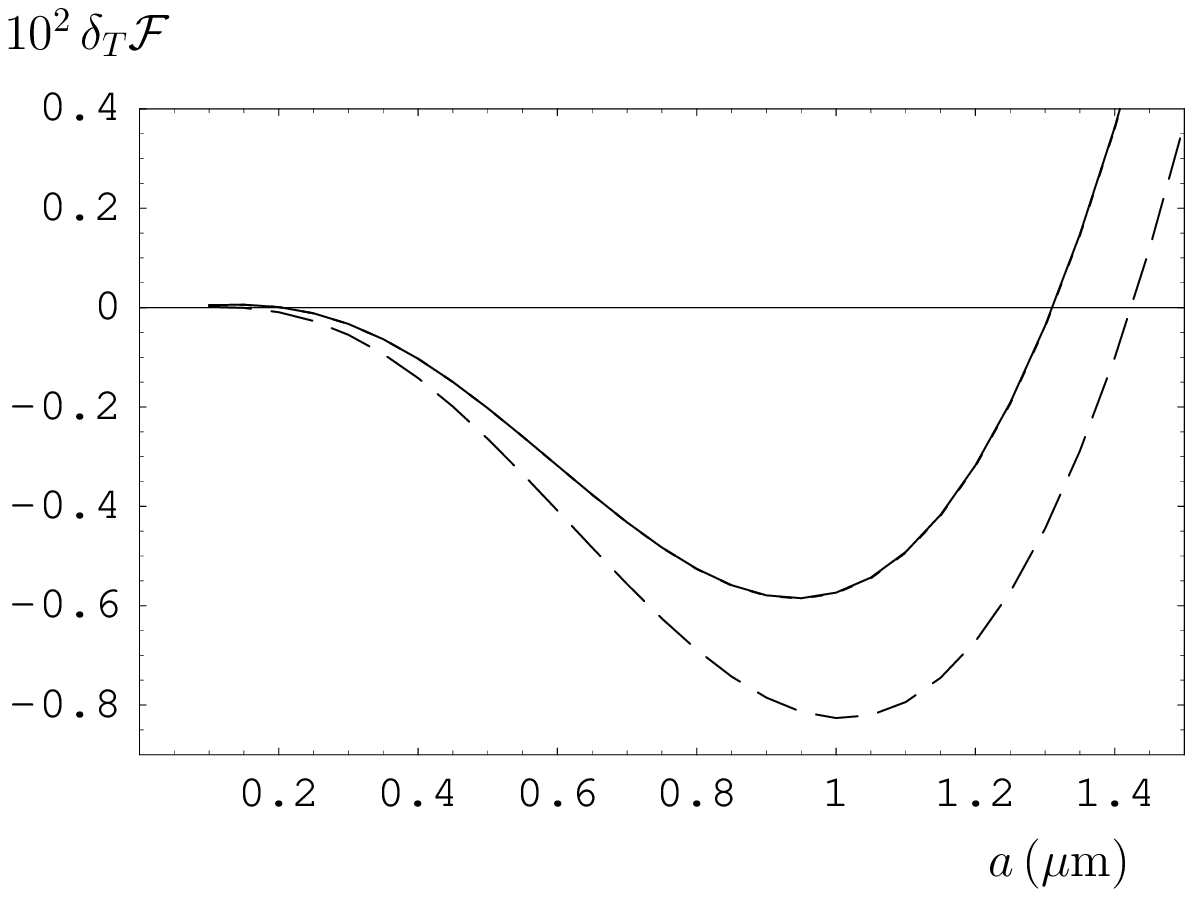}
\vspace*{-9cm}
\caption{
Relative thermal correction to the Casimir energy at 
$T=300\,$K as a function of separation for two plates,
one made of real 
metal (Au) and the other of dielectric (Si)
(solid line).
The same is shown by the dashed line for an ideal metal and 
dielectric with $\varepsilon_0^{\rm D}=11.67$.
}
\end{figure*}
\begin{figure*}
\vspace*{-8cm}
\includegraphics{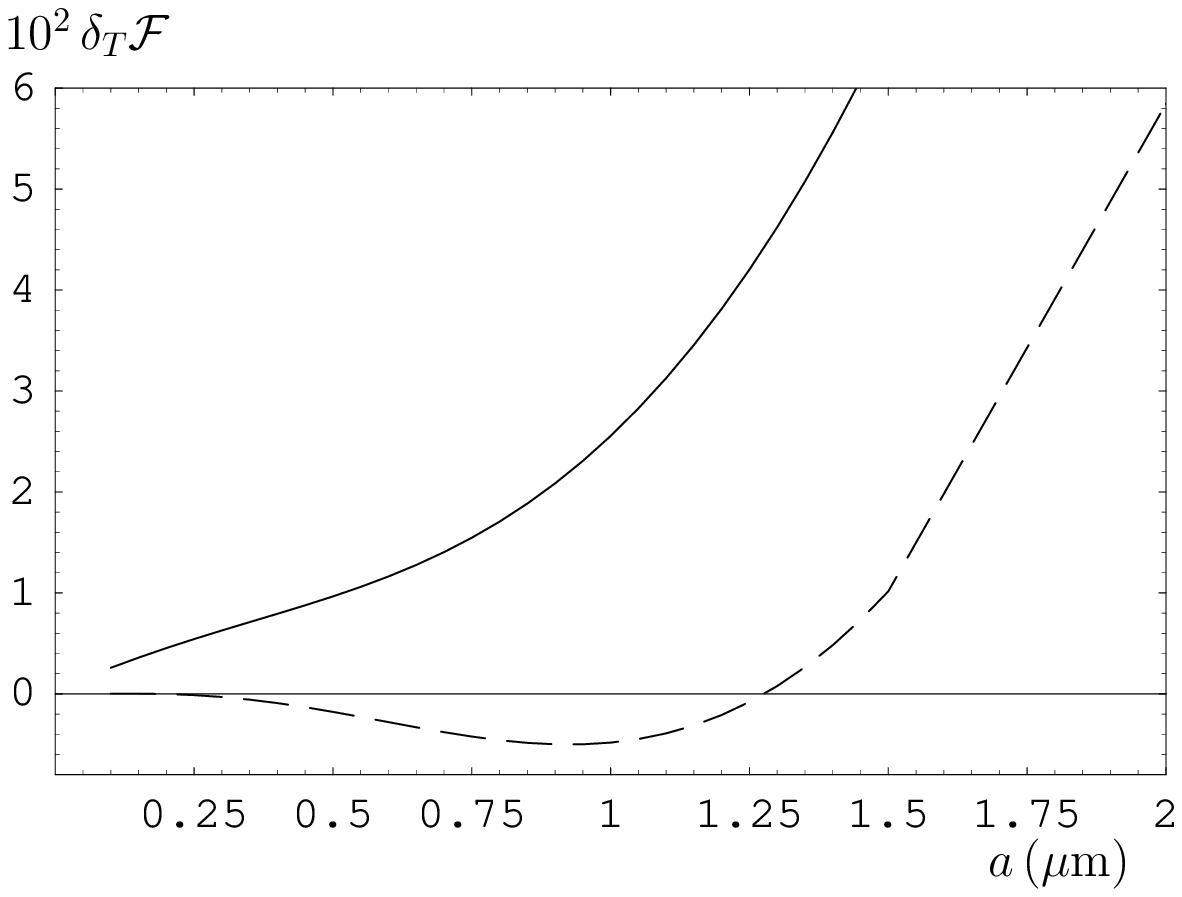}
\vspace*{-9cm}
\caption{
Relative thermal correction to the Casimir energy at
$T=300\,$K as a function of separation for two plates,
one made of real metal (Au) and the other of
dielectric ($\alpha\mbox{-Al}_2\mbox{O}_3$)
(solid line).
The same is shown by the dashed line for an ideal metal and 
dielectric with $\varepsilon_0^{\rm D}=10.1$.
}
\end{figure*}
\end{document}